\documentclass[twocolumn,twoside,slac]{revtex4}
\usepackage{graphicx}
\usepackage{fancyhdr}
\begin{document}

\title{The Persint visualization program for the ATLAS experiment}

%

\author{D. Pomar\`ede}
\affiliation{Commissariat \`a l'Energie Atomique DSM/DAPNIA/SEDI,
CEN Saclay, 91191 Gif-sur-Yvette, France}
\author{M. Virchaux}
\affiliation{Commissariat \`a l'Energie Atomique DSM/DAPNIA/SPP,
CEN Saclay, 91191 Gif-sur-Yvette, France}

\begin{abstract}
The Persint program is designed for the three-dimensional
representation of objects and for the interfacing and access to a variety of
independent applications, in a fully interactive way. Facilities are provided
for the spatial navigation and the definition of the visualization properties,
in order to interactively set the viewing and viewed points, and to obtain the
desired perspective. In parallel, applications may be launched through the use
of dedicated interfaces, such as the interactive reconstruction and display of
physics events. Recent developments have focalized on the interfacing to
the XML ATLAS General Detector Description AGDD, making it a widely used
tool for XML developers. The graphics capabilities of this program were 
exploited in the context of the ATLAS 2002 Muon Testbeam where it was used as
an online event display, integrated in the online software framework and
participating in the commissioning and debug of the detector system.
\end{abstract}

\maketitle

\thispagestyle{fancy}


\section{INTRODUCTION}
Graphics programs in High Energy Physics experiments are necessary to
visualize the detectors geometry and display physics events.
They help in the commissionning phase of the experiment in debugging and
understanding the systems. They must provide an interactive access to the
event data and their reconstruction through online and offline algorithms.
As such they play an important role in the search and discovery of new
physics. 

Persint (Perspectively interacting) is an interactive visualization
program~\cite{persint} which is developed primely for the ATLAS 
collaboration~\cite{atlas}. The
display of objects and the interactivity between the user
and objects or applications is realized through the use of the
HIGZ package~\cite{higz}.
Facilities dedicated to the spatial navigation and the 
definition of the visualization properties are provided, in order
to interactively set the viewing and viewed points, so as obtain the desired 
perspective. In parallel, applications may be launched through the
use of dedicated interfaces, such as the interactive reconstruction
and display of physics events.

\section{DESCRIPTION OF THE PROGRAM}

\subsection{Scope of the program}

Persint was originally developed as a tool for debugging and optimizing
the ATLAS Muon pattern and track reconstruction algorithm
Muonbox~\cite{muonbox1}.
It was early realized that traditional event displays based on simple
projections were not fit to the task of visualizing events in the toroidal,
inhomogeneous field of the muon spectrometer. Only a three-dimensional
interactive tool with navigation capabilities is well suited for such a task. 

The program is now routinely  used for Event Display of hits and
reconstruction objects of complex Monte Carlo events and in TestBeam
applications, both online and offline.
Persint is used for the visualization of detector geometries :
as such it is interfaced to AMDB (the ATLAS Muon Database) and the AGDD-XML
ATLAS Generic detector description.

Other applications include the visualization of the magnetic field,
the Level-1 Muon Trigger Logic, the interactive generation of tracks, and the
display of XML event hits.

\subsection{Features}

The program offers the following features~:
\begin{itemize}
\item 3-dimensional representation of objects in full volumes
or wire frames
\item computation of hidden faces
\item highlight of volumes edges
\item lighting intensity effects on volume facets
\item detection of clashing volumes and highlight of intersections
\item boolean volume operations (addition, subtraction, intersection)
\item spatial navigation with real-time displacements
\item focal length adjustable at will (from isometry to wide-angle)
\item interfaces and access to applications
\item save to postscript
\item documentation
\end{itemize}

\subsection{Design}

The core of the program is written in Fortran90.
Dynamic data objects are extensively exploited in order to minimize
the memory usage which can become important when a large numbers of
volume are displayed, a situation typical to modern HEP experiments,
both with respect to the detector description and the event data
themselves. 

Persint uses the following features of F90 :
\begin{itemize}
\item usage of Modules, with procedure interfaces
\item Polymorphism
\item Recursive functions
\item Dynamic arrays
\item Allocate, for local memory management
\item Pointers
\item String operations
\item Array manipulations
\item Free source forms
\end{itemize}

The graphics interface is HIGZ~\cite{higz} and in general the program
makes extensive use of the CERNLIB~\cite{cernlib}. It is built on 26000
lines on code and is part of the Saclay Muon software suite. As such
it is interfaced directly to the geometrical database Amdbsimrec 
(13000 lines), the magnetic field database Bfielddtb (3500 lines) and
the track reconstruction algorithm Muonbox (58000 lines)~\cite{muonbox1}.

The code is highly modular, organized in patches. The XML section is
such an example of a patch. It contains everything needed to parse,
store transiently and generates volumes from standardized XML files.

The computation of volumes with hidden faces is based on the
analytical computation of facets' edges in space. This computation
also provides the detection of clashes and the highlights of the volumes
intersections, and supports boolean volume operations. 

The program is designed to operate in standalone mode, keeping the
possibility of integration in frameworks. It was for example successfully
used in the context of the online software framework of the ATLAS testbeams.

\subsection{Releases and documentation}

The program is released officially at CERN on AFS in a public 
directory~\footnote{/afs/cern.ch/atlas/offline/external/Persint/}
since version 2.00 for three platforms : Linux, Sun-Solaris Unix and
Compaq/Alpha Unix. This repository contains the source and executable
files together with documentation and useful input databases, and for
developers, the tools necessary to build, compile and load the code.
The latest released version is 3.00.

The documentation can be found in the Persint Manual located 
in the public repository or on the dedicated Persint Web
Page~\footnote{http://cern.ch/Atlas/GROUPS/MUON/persint.html}.
This page provides updates on the releases.

\section{PRESENTATION OF THE PROGRAM}

\subsection{General layout}

\begin{figure}[h]
\centering
\includegraphics[width=82mm]{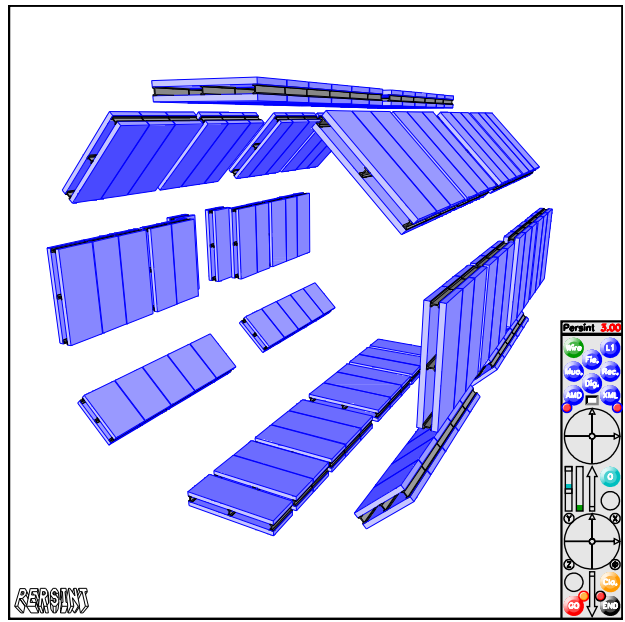}
\caption{Layout of the Persint Display Window, featuring the Navigator interface
and an example of 3D volumes.} \label{layout}
\end{figure}

The program is designed around the command window and the HIGZ display
window. A typical view obtained with the Display Window is shown in
Figure~\ref{layout}.
The command window is used essentially to load files and type
commands interpreted by the code and documented in the 
Persint Manual.
The display window however is organized in
such a way to provide maximum interactivity to avoid typing commands
as much as possible.

\begin{figure*}[t]
\centering
\includegraphics[width=140mm]{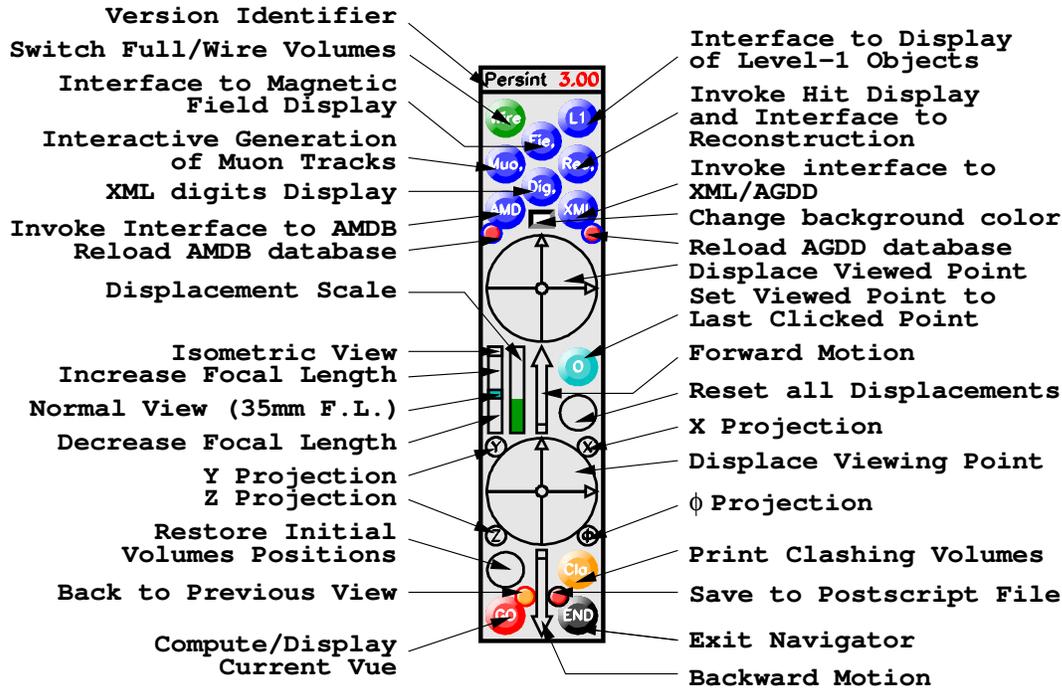}
\caption{Layout of the Navigator.} \label{nav}
\end{figure*}

The display window is dedicated to the colour display of both the
objects computed by the program for visualization and the graphical
interfaces. The primary interface is the Navigator.

The Navigator supports the navigation facilities necessary 
to survey the entire detector space and gives access to the various
graphics applications provided by Persint. Its layout is shown in 
Figure~\ref{nav}. It consists in an ensemble of push buttons and gauges
which can be clicked, so as to modify the visualization properties
or access lower level interfaces.

\subsection{Navigation}

The spatial navigation, as well as the modification of the visualization 
properties, are realized through the use of gauges which define the 
amplitudes of the movements of the Viewing and Viewed Points and of the 
change in focal length of the observer eye, such as to obtain the desired 
perspective. Each action is validated with the ``{\bf GO}" Push Button.

Alternatively, the navigation can be performed through real time
displacements of the
observer point. Continuous displacements are obtained by dragging the
cursor over the display itself, thus bypassing the use of the Navigator.

\subsection{Visualization properties}

\begin{figure*}
\centering
\includegraphics[width=150mm]{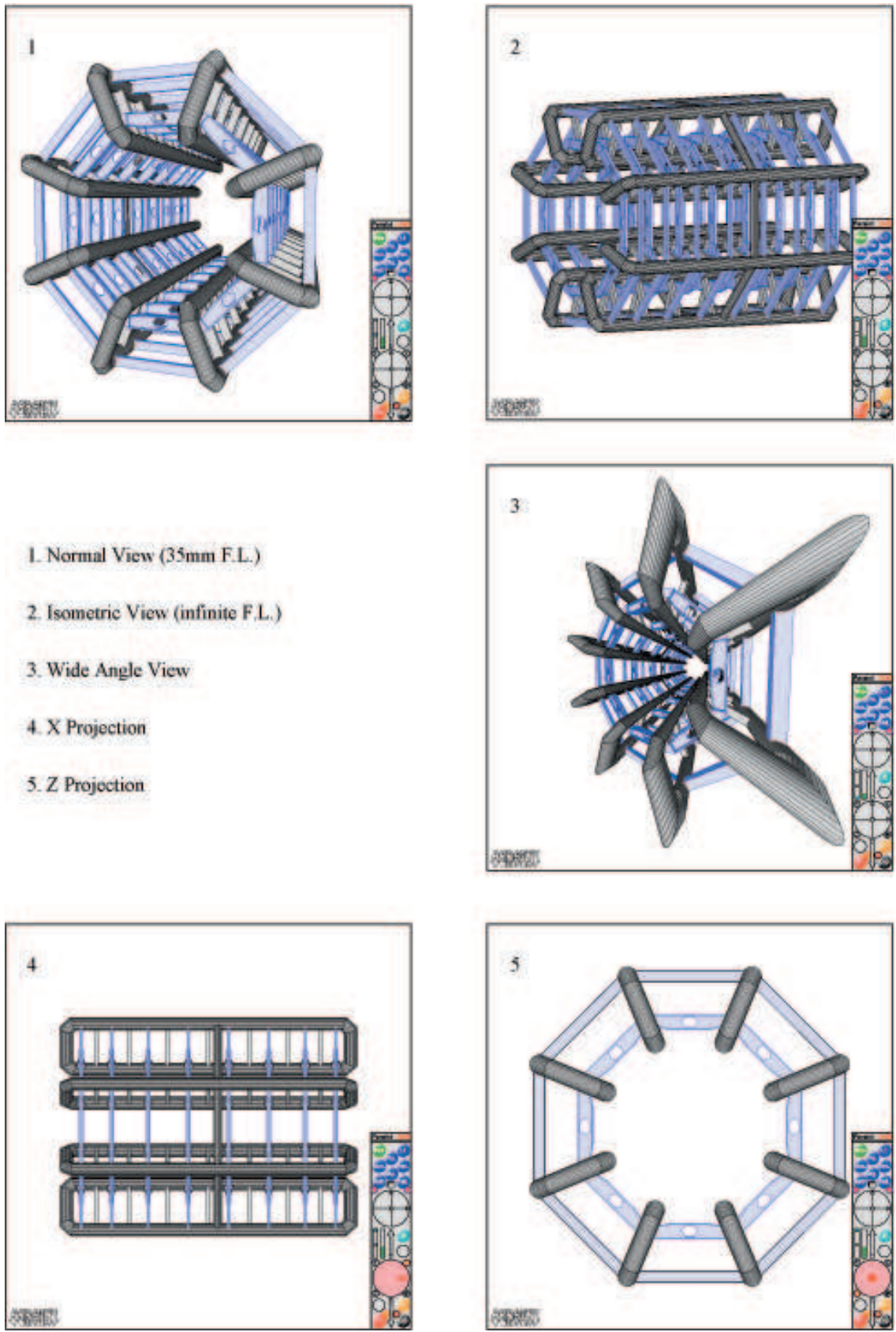}
\caption{Example of perspectives. The ATLAS Muon Spectrometer Barrel Toroid
is displayed under normal viewing condition, isometric view, wide angle 
view and projections. Note the corresponding settings in the Navigator interface.}
\label{expers}
\end{figure*}

Persint offers the possibility to modify interactively the visualization
focal length. In the Navigator, a sectioned gauge may be used to select
the Normal View, equivalent to a focal length of 35 mm,
the Isometric View, equivalent to an infinite focal length, or
to increase or decrease the focal length at will.

Moreover, X, Y, Z, or $\phi$ projections of the volumes on display
may be obtained by clicking dedicated push buttons (see Figure~\ref{nav}).
In the case of  X, Y, and Z projections, the observing point position
(e.g. X$>0$ or X$<0$) can be modified by clicking the ``Displace Viewing
Point" 2D gauge on the corresponding half. In the case of
$\phi$ projection, this same gauge can be used to define the value of
$\phi$.
Examples of visualization of the same volumes with different perspectives 
are presented in Figure~\ref{expers}.

\section{EVENT DISPLAY}

\begin{figure*}
\centering
\includegraphics[width=162mm]{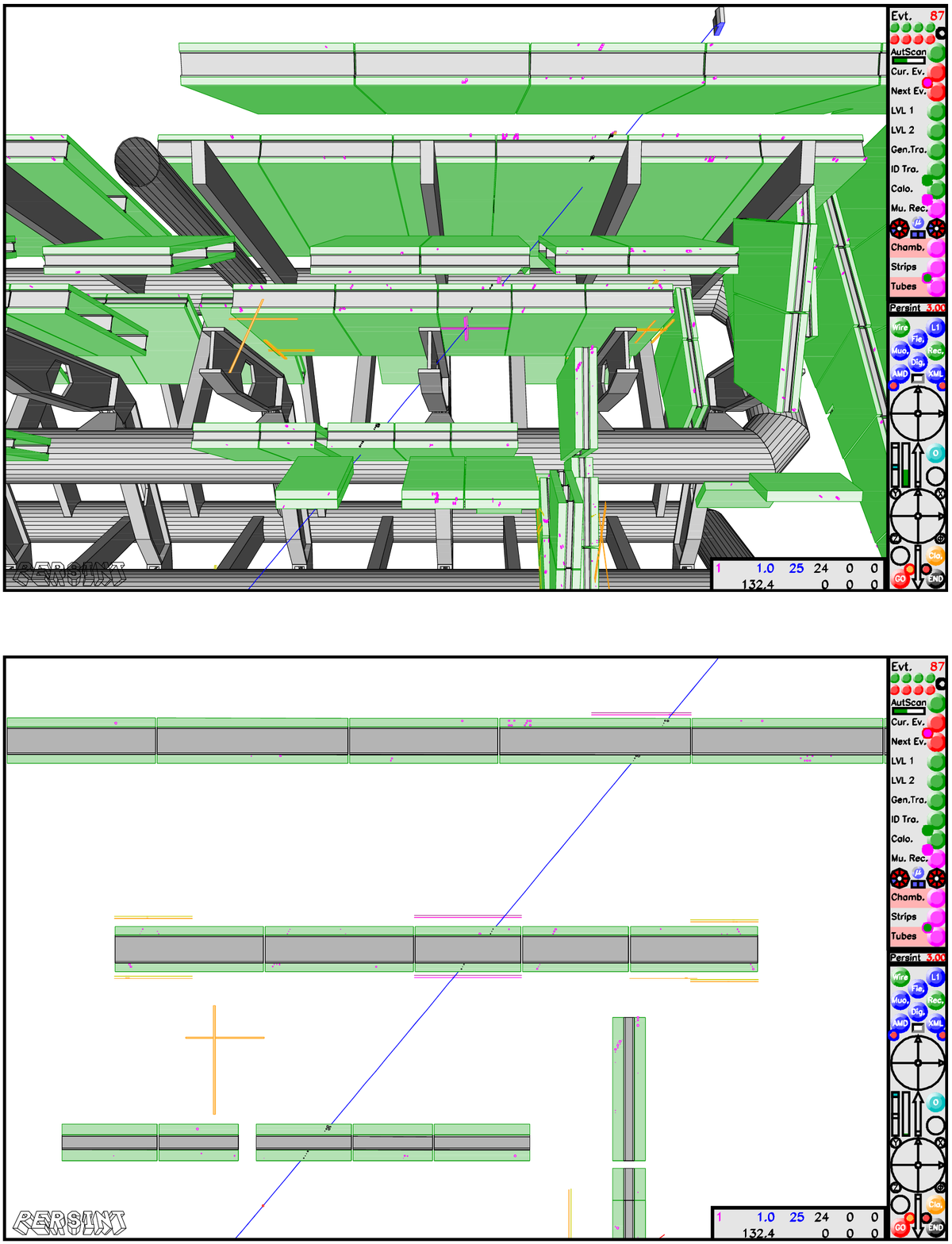}
\caption{Event Display. These views show the display of the same Monte
Carlo event, a single muon track with high luminosity pile-up, 
generated with a safety factor 5 on cavern component. The blue line corresponds to
a candidate track reconstructed by Muonbox. The top view is a 3-dimensional
representation, the bottom view is an isometric projection.
The Interactive Reconstruction and Event Display Interface is visible on
top of the Navigator, as well as the event statistics window.}
\label{evd}
\end{figure*}

The interactive reconstruction and event display (EVD) application of Persint
is dedicated to the
reading of event hit files, the visualization of these hits, and the 
reconstruction of the event through coupling to the Muonbox
program~\cite{muonbox1}.

The three-dimensional view of complex events is used to understand complex
events and debug the reconstruction algorithms. Figure~\ref{evd} shows
a display of a Monte Carlo event in the ATLAS Muon spectrometer in the context
of a high background environment. Such events are generated with the Geant3-based 
DICE simulation~\cite{dice}.
All hits and activated muon chambers are
displayed. The reconstructed track is displayed in blue. Visual inspection
confirms the efficiency of the reconstruction algorithm in terms of pattern
recognition and fitting.

\begin{figure}[h]
\centering
\includegraphics[width=83mm]{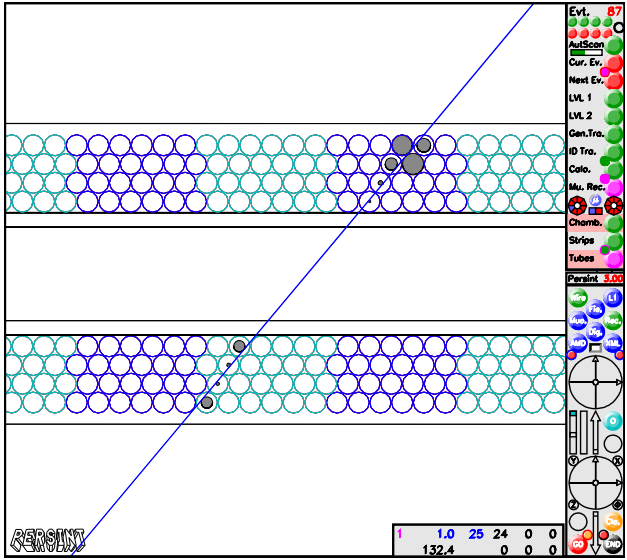}
\caption{Event Display resulting from a zoom on a muon chamber.
Visual investigation of such closeups allows to check the
consistency of the event
properties versus the reconstruction output.}
\label{evdDCbil}
\end{figure}

The projective views allows to further examine the events, though it is
necessary to restrict the investigation to a limited region of the apparatus.
Zooming on detectors is performed to very carefully check the event
properties and reconstruction consistency, as illustrated in 
Figure~\ref{evdDCbil}.

In Figure~\ref{evd} one notes the Interactive Reconstruction and Event Display 
Interface opened on top of the Navigator. This interface is equipped with an
ensemble of push-buttons to
set the visual properties of the displayed event and to activate the
reconstruction. A separate embedded window is used to display the statistics
of the event. The statistics consist essentially  in the number of hits of
the event, the number of hits taken into account in the candidate tracks, the
reconstructed momentum, and if required from the datacards in case of physics
channels analysis, the reconstructed invariant mass.

The interfacing between Persint and Muonbox makes it possible
to act interactively on 
the event characteristics to understand which role they play. For instance,
a hit tube or a strip can be removed interactively from the event. Then,
when re-running the reconstruction algorithms, this removed tube will not
be taken into account.

The Persint EVD was used in the design and optimization phase of
the Muon Spectrometer, participating in the performance studies of the apparatus
and reconstruction software~\cite{muontdr}. It then contributed in refining the
performances of this software on complex Monte Carlo physics
channels~\cite{phystdr}.

\begin{figure}[h]
\centering
\includegraphics[width=83mm]{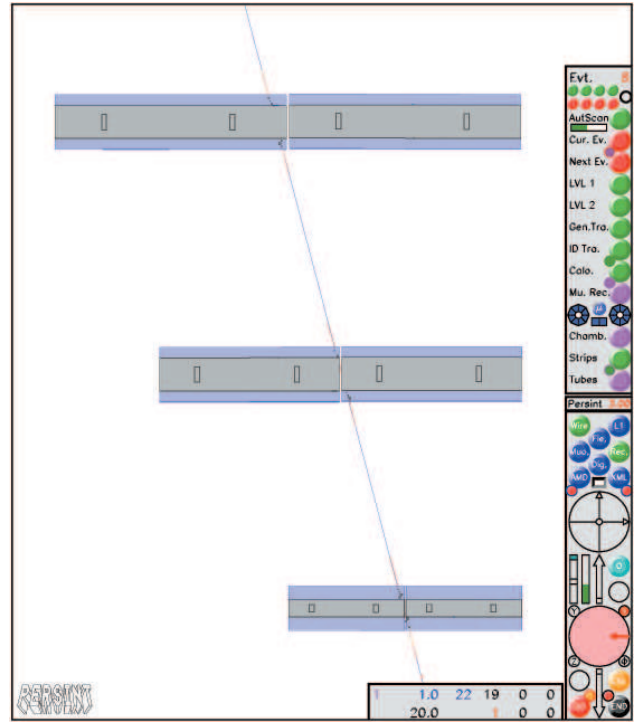}
\caption{H8 2002 Muon Testbeam Event Display.}
\label{tb}
\end{figure}

The program is used in the context of muon testbeams to display events
on either online of offline mode~\cite{h8_2001}. Event displays are
useful to testbeams
in order to validate the commissioning of the entire chain of detectors
and data acquisition. When data is recorded in a routine mode, the EVD
participate in the online monitoring and the quality control of the dataflow.
An example of testbeam event is shown in Figure~\ref{tb}. It consists
in a staight muon track reconstructed in the Barrel stand of the H8 2002
Muon System tests~\cite{h8_2002}. Chambers misalignment is corrected using
conditions alignment data.

\begin{figure*}[t]
\centering
\includegraphics[width=170mm]{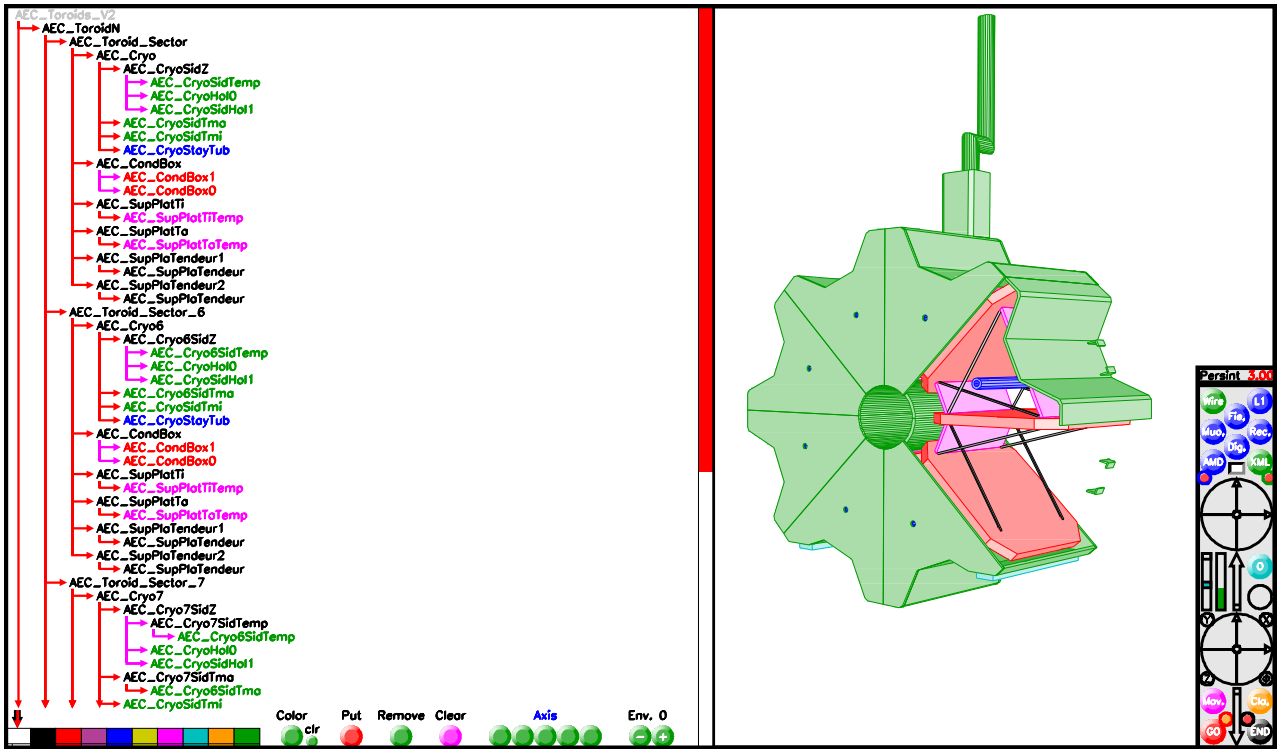}
\caption{Visualization of AGDD-XML volumes and associated XML tree structure.
In this example, one of the Endcap Toroid of the Atlas Muon spectrometer 
is displayed on the right. The corresponding XML tree is shown on the left.}
\label{xml}
\end{figure*}

\begin{figure}[ht]
\centering
\includegraphics[width=82mm]{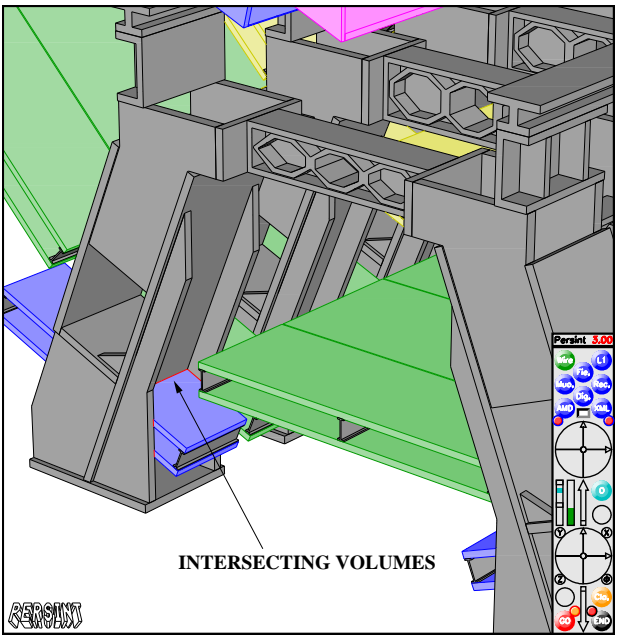}
\caption{Visualization of AMDB volumes. Intersection of volumes are 
detected and highlighted in red.}
\label{clash}
\end{figure}

\section{GEOMETRY VISUALIZATION}

Powerful graphics tool are necessary to develop and debug the geometrical 
descriptions of the complex HEP detectors. Not only 3D viewers and projective
capabilities must be used, but also the possibility to automatically detect
geometrical conflicts.

An essential feature of geometry visualization programs is to be able to
delineate precisely the volume edges. This is of primary importance in order
to clearly identify the volumes boundaries in space and their possible
interferences. Such visual property offers a much better understanding and
comprehension of the detector geometry to the end user.

Modern geometrical description in HEP makes extensive use of boolean volume
operations, which greatly simplifies the task of describing complicated 
geometries. Visualization tools must therefore support such technology.

The Persint program is interfaced to the two detector description used for
the ATLAS Muon Spectrometer : AGDD-XML and AMDB.

\subsection{Interface to AGDD-XML}

The Atlas Generic Detector Description is based on standardized XML files.
Such a description is used to describe the Muon Spectrometer
geometry~\cite{agdd}. This description is implemented in the reconstruction
algorithm.

The Persint program benefits from an XML parser to interpret the corresponding
geometrical description. 
The working of the interface to AGDD is based on the representation of a fully 
clickable tree-like structure allowing selection of the desired volumes
for display. An operating example is given in Figure~\ref{xml}, showing
both the XML structure and the corresponding volumes, in this case one of the
ATLAS Endcap toroid.

The interface offers the representation of the volume tree in its full
depth. The user can deploy or hide sub-structures by clicking
Volume Tree Pointers appearing as arrows.

Conveniently, geometries can be visualized in parallel as they are being
developped. After a possible modification, clicking on a a dedicated
push button of the Navigator (see Figure~\ref{nav}) results in the reloading
of the database.

\subsection{Interface to AMDB}

The ATLAS Muon Database is an object-oriented database for simulation
and reconstruction~\footnote{see http://cern.ch/muondoc/software/Database/}.
It is extensively used in the detectors and physics performance studies 
conducted since the origin of ATLAS~\cite{muontdr,phystdr}. In particular,
it is used synchroneously by the DICE Geant3 simulation of the
experiment~\cite{dice} and the muon reconstruction program 
Muonbox~\cite{muonbox1}. 

This database is heavily used to study the ongoing
changes to the spectrometer layout~\cite{crack}.
Such changes are caused by the need to
take into account integration and access constraints. Also, the possible
staging of ATLAS will lead to multiple geometrical configuration, to be
optimized by simulation. In this context of evolving geometries, Persint
provides a tool to carefully scrutinize and validate the geometrical
databases. One of the important capabilities of Persint is to detect 
and highlight intersecting or clashing volumes. An illustration is given
in Figure~\ref{clash}, where a conflict between a muon chamber and the
feet of the detector is observed.

\section{BOOLEAN OPERATIONS}

Persint offers the rare feature of being able to compute fully and
display volumes resulting from Boolean operations.

\begin{figure}[t]
\centering
\includegraphics[width=82mm]{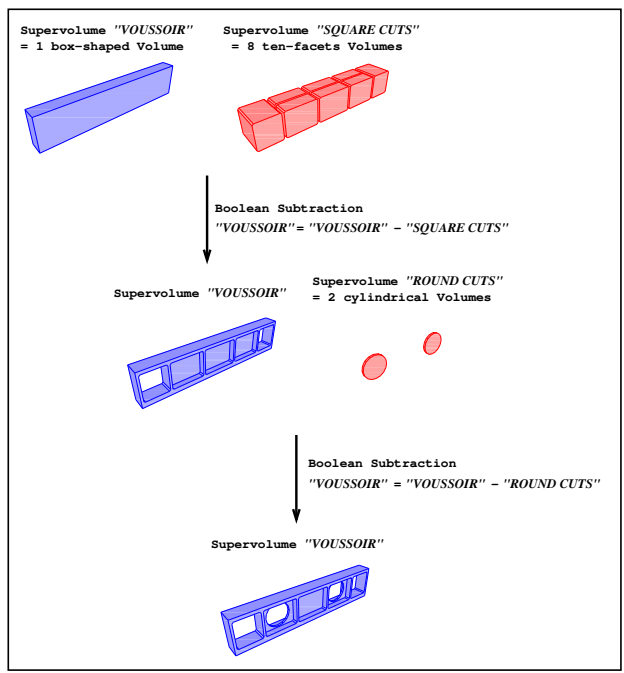}
\caption{Example of Boolean Volume Operations.}
\label{labelvouss}
\end{figure}

Being given two initial Supervolumes (a Supervolume is an ensemble 
of volumes), 
the program is able to perform the three basic operations
(addition, subtraction, intersection) to generate
a new Supervolume. This facility is independent of the complexity of
the initial objects.
An example is shown in Figure~\ref{labelvouss}, which illustrates
how the program can generate the voussoirs of the Barrel Toroid.

Using this feature of Persint, it is possible to generate 
complex-shaped volumes from very basic objects.
The program provides, through the Command Mode, a set of instructions
allowing to generate basic volumes and to perform Boolean operations.

\section{OTHER APPLICATIONS}

The Persint program is interfaced to a variety of application. They are all
accessed through the Navigator (see Figure~\ref{nav}).

The Level 1 Interface is the application allowing the interactive access 
to the Level-1 Muon Trigger configuration datafiles and the visual
representation of the variety of objects involved in the trigger
decision chain. This interface is extensively described in~\cite{L1persint}.

\begin{figure}[t]
\centering
\includegraphics[width=82mm]{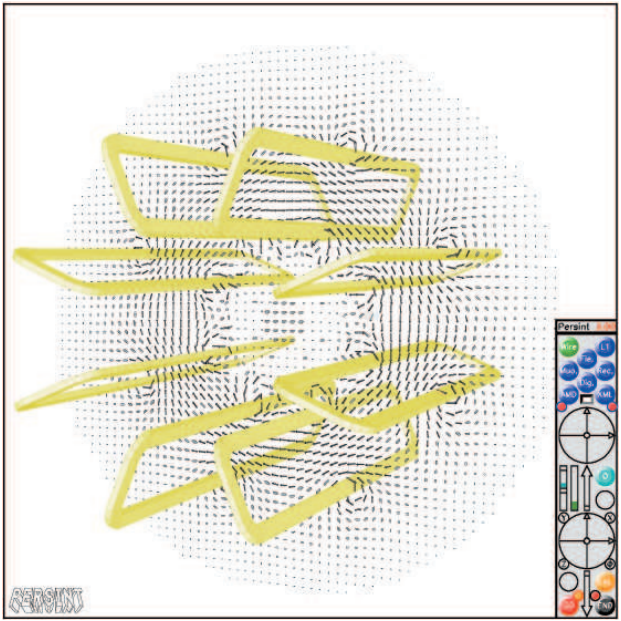}
\caption{Visualization of the magnetic field and magnet elements.}
\label{magfield}
\end{figure}

The Magnetic Field Interface provides a fully interactive 3D visualization of 
the Magnetic Field and the magnet elements. This application is coupled to
the Bfielddtb set of 
subroutines that read the field database and compute the field at
any given point~\footnote{see http://cern.ch/Atlas/GROUPS/MUON/magfield/}.
The field is visualized by means of arrows located at a series of lattice points. 
An example is shown in Figure~\ref{magfield}. 
The direction and length of the arrow correspond to the direction and magnitude
of the B-field. The lattice, either 2D or 3D,
is customized by the user in terms of mesh and spatial limits, allowing
to build the vector field either in very localized regions or over large
distances.

The Muon Tracks Interface is used to generate and compute tracks of muons
throughout the entire detector system. It is coupled to Muonbox, to the detector
description and the magnetic field database, in order to compute the track's
trajectory, taking into account the deflections in the field, the multiple scattering,
and the energy loss in matter.

\section{PERFORMANCES}

Performances are difficult to estimate, as the visualization programs
have too face many possible situations in terms of user demand, and number
of volumes to display.

A test is performed  here as follows : the display and navigation
capability is estimated on a 2.8 GHz Pentium 4. It requires 0.65 s
to compute and display the 2300 voulumes with 15000 facets of the
Muon precision chambers system.
The real-time displacements are fluid ($>$~10 views per second) for
a number of volumes $<$~400.

An event scan on testbeam data provides another performance measurement.
In the exercise consisting in the display of events in loop with 6 chambers
and typically 20 hits, it needs 0.02 s/event for bare EVD. This an order of
magnitude less than the time needed to reconstruct the event : 0.20 s/event.

\section{CONCLUSIONS AND PROSPECTS}

The present design of the program has strong advantages. The user can
download a single executable binary files. No special dynamic library loading,
no fancy APIs need to be installed. The program is efficient even if
operated remotely, e.g. on remote clusters from a X-terminal.

On the other side, the program has shortcomings. It uses a limited graphics interface 
(256 colors in the current version of HIGZ, possibly extended to 1024). It
does not profit from the high performance capabilities of graphics cards when operated
locally on PCs (Z-buffer). 

In the developments under consideration, a migration to OpenGL is envisaged. However
some of the features of the present design would be lost. For example, such technology
based on Z-buffer doesn't allow to delineate the volume edges. This property is crucial
for HEP detectors.

It is considered to integrate parts of Persint in the ROOT data analysis 
framework. The graphics
capabilities of ROOT would be enhanced by using the calculus of hidden faces and
intersecting volumes. In parallel, the integration of Persint in the ATHENA Atlas
software framework for reconstruction and analysis can be studied.

The utilization of Persint will continue as an event and geometry display for the ATLAS
experiment. It will be used as online EVD in the H8-2003 Muon Testbeam, and in the
H8-2004 ATLAS Combined Testbeam. It will have an important role to play
in the commissioning of ATLAS, participating in the debug of the detector (cosmics
run, calibration runs,...). Finally it is a candidate for the ATLAS Online EVD
for monitoring of data-taking.

\end{document}